\documentclass[%
 aip,
 amsmath,amssymb,
 reprint,%
]{revtex4-1}

\usepackage{graphicx}
\usepackage{dcolumn}
\usepackage{bm}

\usepackage[utf8]{inputenc}
\usepackage[T1]{fontenc}
\usepackage{mathptmx}
\usepackage{etoolbox}
\usepackage{subfigure}
\usepackage{xcolor}

\makeatletter
\def\@email#1#2{%
 \endgroup
 \patchcmd{\titleblock@produce}
  {\frontmatter@RRAPformat}
  {\frontmatter@RRAPformat{\produce@RRAP{*#1\href{mailto:#2}{#2}}}\frontmatter@RRAPformat}
  {}{}
}%
\makeatother
\begin{document}

\preprint{AIP/123-QED}

\title[]{Calculation of toroidal Alfvén eigenmode mode structure in general axisymmetric toroidal geometry}
\author{Guangyu Wei}
\affiliation{Institute for Fusion Theory and Simulation and Department of Physics, Zhejiang University, Hangzhou 310027, China}
\affiliation{Center for Nonlinear Plasma Science and C.R. ENEA Frascati, Via E. Fermi 45, Frascati, Italy}
\author{Matteo Valerio Falessi}
\affiliation{Center for Nonlinear Plasma Science and C.R. ENEA Frascati, Via E. Fermi 45, Frascati, Italy}
\affiliation{Istituto Nazionale di Fisica Nucleare (INFN), Sezione di Roma, Piazzale Aldo Moro 2, Roma, Italy}
\author{Tao Wang}
\affiliation{Institute for Fusion Theory and Simulation and Department of Physics, Zhejiang University, Hangzhou 310027, China}
\affiliation{Center for Nonlinear Plasma Science and C.R. ENEA Frascati, Via E. Fermi 45, Frascati, Italy}
\author{Fulvio Zonca}\email{fulvio.zonca@enea.it}
\affiliation{Center for Nonlinear Plasma Science and C.R. ENEA Frascati, Via E. Fermi 45, Frascati, Italy}
\affiliation{Institute for Fusion Theory and Simulation and Department of Physics, Zhejiang University, Hangzhou 310027, China}
\author{Zhiyong Qiu}
\affiliation{Key Laboratory of Frontier Physics in Controlled Nuclear Fusion and Institute of Plasma Physics, Chinese Academy of Sciences, Hefei 230031, China}
\affiliation{Center for Nonlinear Plasma Science and C.R. ENEA Frascati, Via E. Fermi 45, Frascati, Italy}

\date{\today}

\begin{abstract}
A workflow is developed based on the ideal MHD model to investigate the linear physics of various Alfvén eigenmodes in general axisymmetric toroidal geometry, by solving the coupled shear Alfvén wave (SAW) and ion sound wave (ISW) equations in ballooning space. The model equations are solved by the FALCON code in the singular layer, and the corresponding solutions are then taken as the boundary conditions for calculating parallel mode structures in the whole ballooning space. As an application of the code, the frequencies and mode structures of toroidal Alfvén eigenmode (TAE) are calculated in the reference equilibria of the Divertor Tokamak Test facility (DTT) with positive and negative triangularities, respectively. By properly handling the boundary conditions, we demonstrate finite TAE damping due to coupling with the local acoustic continuum, and find that the damping rate is small for typical plasma parameters. 
\end{abstract}
\pacs{}

\maketitle

\section{Introduction}
In magnetically confined fusion plasmas, excitation of shear Alfvén wave (SAW) instabilities by energetic particles (EPs) via wave-particle resonance has been one of the major concerns due to their potential role in EP transport \cite{LChenRMP2016}. Therefore, a comprehensive understanding of the dynamics of SAW fluctuations in fusion plasmas is of crucial importance. It is well known that, in toroidal plasma devices such as tokamaks, SAWs and ion sound waves (ISWs) are characterized by continuously varying continuous spectra. The periodic variation of equilibrium quantities along the magnetic field lines leads to the breakup of the continuous spectrum and formation of SAW-ISW frequency gaps \cite{CZChengAP1985, CZChengPoF1986}, where discrete eigenmodes may exist. The discrete Alfvén eigenmodes (AEs), such as toroidal Alfvén eigenmode (TAE) \cite{CZChengAP1985,CZChengPoF1986, WHeidbrinkPRL1991} and beta-induced Alfvén eigenmode (BAE) \cite{WHeidbrinkPRL1993}, which are inside the frequency gaps and are free of significant continuum damping \cite{MRosenbluthPRL1992,FZoncaPRL1992,FZoncaPoFB1993}, can be easily driven unstable by EPs with a low threshold condition and, thus, have attracted a lot of attention in fusion research. If the EP drive is sufficiently strong to overcome the continuum damping, it can give rise to another type of discrete fluctuations known as energetic particle modes (EPMs) \cite{LChenPoP1994} originated from the SAW continuous spectrum. Therefore, it is particularly important to precisely calculate the structures of SAW continuous spectrum and the frequencies of various AEs in realistic magnetic configurations with plasma non-uniformities included, which can serve as a valuable reference for analyzing and interpreting simulation results and experimental observations.

The coupled SAW-ISW continuous spectrum can be obtained by solving the equations that describe the fluctuations propagation along the equilibrium magnetic field lines \cite{MSChuPoF1992, MFalessiPoP2019_continuum, MFalessiJPP2020}. This is equivalent to the typical approach that takes a spectral decomposition of the fluctuations in poloidal and toroidal angles and computes the null space (kernel) of the matrix of the highest order radial derivative \cite{CZChengPoF1986}, corresponding to radially singular SAW-ISW structures. It is worthy noting that, the former method is convenient when dealing with high toroidal mode numbers, and can be straightforwardly extended to kinetic description\cite{MFalessiPoP2019_continuum}. Besides, by solving the differential equations of SAW-ISW continuous spectrum in ballooning space, the obtained solutions also determine the asymptotic behaviors of the discrete AEs in the frequency gaps \cite{LChenPoP2017, MFalessiPoP2019_continuum}.

In previous works\cite{MFalessiPoP2019_continuum,MFalessiJPP2020,GPucellaPPCF2022}, some of the authors developed the Floquet Alfvén continuum code (FALCON) to calculate the coupled SAW-ISW continuous spectrum in general axisymmetric toroidal geometry. FALCON, along with its associated library called EQUIlibrium Postrpocessing codE (EQUIPE), are open source tools avaible to the fusion community. They have been collaboratively developed\footnote{The access to the code repository can be obtained by writing to the code maintainer at matteo.falessi@enea.it}, validated\cite{GPucellaPPCF2022,LPiron2023} and verified in the last years. In this work, we are dedicated to develop a workflow for the calculation of AEs within this framework and adopt the ideal MHD limit to illustrate the general methodology in a simple case. The extension of present calculation to kinetic description is straightforward thanks to the adoption of ballooning mode representation \cite{JConnorPRL1978, ZLuPoP2012}, and will be pursued in future work. The local dispersion curves calculated by FALCON\cite{MFalessiPoP2019_continuum,GPucellaPPCF2022} can be used to construct the global continuous spectrum. Meanwhile, they can be taken as the boundary condition for calculating the parallel mode structures of discrete gap modes. Therefore, we have developed a general workflow, based on the FALCON code, to calculate the frequencies and mode structures of AEs. As an application of the code, in this work, we calculate the TAE \cite{CZChengAP1985, CZChengPoF1986} spectrum in one of the reference equilibria of the Divertor Tokamak Test facility (DTT)\cite{RAlbaneseFED2017}, and demonstrate the existence of a small but finite TAE damping rate in the ideal MHD limit. This damping results from the coupling between TAE and the acoustic continuum, so it is generally much smaller than the conventional SAW continuum damping for typical plasma parameters, due to the different polarizations of the eigenmode and the local ISW continuum. The dependence of this effect on plasma shaping is illustrated here as we calculate TAE in a DTT equilibrium with negative triangularity (NT), demonstrating the general applicability of the workflow across different geometries. We find minimal deviation from previous results with positive triangularity (PT), which is consistent with the experimental observations on DIII-D tokamak\cite{MVanZeelandNF2019}.

The paper is organized as follows. In Sec. \ref{Theoretical Framework}, we introduce the model equations and numerical approach adopted in the code. In Sec. \ref{Numerical results}, we apply the code to analyze the frequencies and parallel mode structure of TAE in DTT equilibria with positive and negative triangularities and discuss the weak effect of triangularity on TAE. In Sec. \ref{Summary and prospect}, we give a brief summary of current work and outline the future prospects for code development.


\section{Theoretical Framework}\label{Theoretical Framework}

\subsection{Model equations}\label{Model equations}
Following Ref. \citenum{LChenPoP2017}, we introduce our model equations that describe the SAW and ISW fluctuations coupled by the magnetic curvature, which consist of the vorticity equation 
\begin{equation}\label{SAW}
  \begin{aligned}
    &B_0\nabla_\parallel\left(\frac{1}{B_0}\nabla_\perp^2\nabla_\parallel\Phi_s\right)-\bm{\nabla}_\perp\cdot\left(\frac{\partial_t^2}{v_A^2}\bm{\nabla}_\perp\Phi_s\right)\\
    &+\frac{8\pi}{B_0}(\bm{b}_0\times\bm{\kappa})\times\bm{\nabla}_\perp\left[\frac{\bm{b}_0}{B_0}\cdot(\bm{\nabla}P_0\times\bm{\nabla}_\perp\Phi_s)\right]\\
    &+\frac{4\pi}{cB_0}\left[\bm{b}_0\times\bm{\nabla}_\perp(\nabla_\parallel\Phi_s)\right]\cdot\bm{\nabla}J_{0\parallel}\\
    &=-\frac{8\pi}{cB_0}(\bm{b}_0\times\bm{\kappa})\cdot\bm{\nabla}_\perp\delta P_\mathrm{comp},
  \end{aligned}
\end{equation}
and the parallel force balance equation
\begin{equation}\label{ISW}
  \begin{aligned}
    &\partial_t^2\delta P_\mathrm{comp}-c_S^2B_0\nabla_\parallel\left(\frac{1}{B_0}\nabla_\parallel\delta P_\mathrm{comp}\right)\\
    &=-\frac{2\Gamma P_0c}{B_0}(\bm{b}_0\times\bm{\kappa})\cdot\bm{\nabla}_\perp\partial_t^2\Phi_s,
  \end{aligned}
\end{equation}
where $\Phi_s$ is the perturbed stream function, related with the perpendicular plasma displacement $\delta\bm{\xi}_\perp=(c/B_0)\bm{b}_0\times{\bm{\nabla}\Phi_s}$, $\delta P_\mathrm{comp}=-\Gamma P_0(\bm{\nabla}\cdot\bm{\xi})$ is the compressional component of the perturbed pressure, $\bm{\kappa}=\bm{b}_0\cdot\bm{\nabla}\bm{b}_0$ is the magnetic curvature, $v_A^2=B_0^2/(4\pi\rho_0)$ and $c_S^2=\Gamma P_0/\rho_0$ are Alfvén and sound speed, respectively, and other notations are standard. In Eq. \eqref{SAW}, the first three terms on the left-hand side represent field line bending, plasma inertia and curvature-pressure coupling, respectively, while the fourth term, representing the well-known kink drive, will be neglected in the following study, since it is typically small for high mode number AEs characterizing DTT plasmas as expected for reactor relevant conditions\cite{TWangPoP2018,TWangPoP2019}. The right-hand side is the contribution of curvature coupling with plasma compressibility. In Eq. \eqref{ISW}, the left-hand side can be regarded as the linear response of ISW in a straight equilibrium magnetic field, and the right-hand side is the correction of finite magnetic curvature.

In this work, we consider a general tokamak geometry with the equilibrium magnetic field expressed by 
\begin{equation}
  \bm{B}_0 = F(\psi)\bm{\nabla}\varphi+\bm{\nabla}\varphi\times\bm{\nabla}\psi,
\end{equation}
where $\psi$ is the poloidal magnetic flux function, and $\varphi$ is the physical toroidal angle. Following our previous works\cite{MFalessiPoP2019_continuum}, we make use of Boozer coordinates $(r, \theta, \zeta)$, where $r(\psi)$ is a radial-like flux coordinate, poloidal angle $\theta$ is chosen such that $\mathcal{J}B_0^2$ depends only on $\psi$, with $\mathcal{J}=(\bm{\nabla}\psi\times\bm{\nabla}\theta\cdot\bm{\nabla}\zeta)^{-1}$ being the Jacobian of the coordinates, and the generalized toroidal angle $\zeta$ is defined such that the magnetic field lines are straight on a flux surface, i.e., 
\begin{equation}
  q=\frac{\bm{B}_0\cdot\bm{\nabla}\zeta}{\bm{B}_0\cdot\bm{\nabla}\theta}
\end{equation}
is a flux function. Focusing on modes with high toroidal mode number and finite magnetic shear, we adopt the ballooning mode representation \cite{JConnorPRL1978, ZLuPoP2012}, where an arbitrary fluctuation field can be decomposed as 
\begin{equation}\label{ballooning_representation}
  \begin{aligned}
    f(r,\theta,\zeta)
    &=\sum_{m\in\mathbb{Z}}A_n(r)e^{i(n\zeta-m\theta)}\int d\vartheta e^{i(m-nq)\vartheta}\hat{f}_{n}(\vartheta, r)\\
    &=2\pi A_n(r)\sum_{\ell\in\mathbb{Z}}e^{in\zeta-inq(\theta-2\pi\ell)}\hat{f}_n(\theta-2\pi\ell, r),
  \end{aligned}
\end{equation}
and the time dependence of the fluctuation field ($\propto e^{-i\omega t}$) is left implicit for the sake of simplicity in notation. On the right hand side of Eq. \eqref{ballooning_representation}, $A(r)$ is the radial envelope of the mode, and $\hat{f}_n(\vartheta, r)$ is the corresponding parallel mode structure in the ballooning space, with $\vartheta$ being the "extended poloidal angle" along the magnetic field line. By writing Eq. \eqref{ballooning_representation}, we have assumed the scale separation between radial envelope and single poloidal harmonics, which enables us to reduce the original two-dimensional problem in the poloidal plane to two one-dimensional problems \cite{FZoncaPoFB1993}, i.e., solving for the parallel mode structure $\hat{f}_n(\vartheta, r)$ in ballooning space at the lowest order and for the radial envelope $A(r)$ at the next step. In this work, we focus on the solution of the parallel mode structure.

Applying the ballooning mode representation and ignoring the effect of radial envelope variation, the coupled SAW-ISW equations Eq. \eqref{SAW} and Eq. \eqref{ISW} become \cite{MFalessiPoP2019_continuum}
\begin{widetext}
  \begin{subequations}\label{SAW-ISW}
    \begin{equation}\label{SAW-ISW-a}
      \begin{aligned}
        & \left[\partial_{\vartheta}^{2}-\frac{\partial_{\vartheta}^{2}\hat{\kappa}_{\perp}}{\hat{\kappa}_{\perp}}+\frac{\omega^{2}\mathcal{J}^{2}B_{0}^{2}}{v_{A}^{2}}
        -8\pi\mathcal{J}^{2}\frac{rB_{0}P_0'}{q\hat{\kappa}_{\perp}\psi'}
        \left(\kappa_{g}\frac{\bm{\nabla}\psi\cdot\hat{\bm{\kappa}}_{\perp}}{\hat{\kappa}_{\perp}|\bm{\nabla}\psi|}-\kappa_{n}\frac{rB_{0}}{q\hat{\kappa}_{\perp}|\bm{\nabla}\psi|}\right)\right]g_{1}(\vartheta)\\
      & =-\left(2\Gamma\beta\right)^{1/2}\frac{\mathcal{J}^{2}B_{0}^{2}}{qR_{0}}\left(\kappa_{g}\frac{\bm{\nabla}\psi\cdot\hat{\bm{\kappa}}_{\perp}}{\hat{\kappa}_{\perp}|\bm{\nabla}\psi|}-\kappa_{n}\frac{rB_{0}}{q\hat{\kappa}_{\perp}|\bm{\nabla}\psi|}\right)g_{2}(\vartheta),\\
      \end{aligned}
    \end{equation}
    \begin{equation}\label{SAW-ISW-b}
      \left(\frac{\mathcal{J}^{2}B_{0}^{2}}{q^{2}R_{0}^{2}}+\frac{\omega_{S}^{2}}{\omega^{2}}\partial_{\vartheta}^{2}\right)g_{2}(\vartheta)
      =-\left(2\Gamma\beta\right)^{1/2}\frac{\mathcal{J}^{2}B_{0}^{2}}{qR_{0}}\left(\kappa_{g}\frac{\bm{\nabla}\psi\cdot\hat{\bm{\kappa}}_{\perp}}{\hat{\kappa}_{\perp}|\bm{\nabla}\psi|}-\kappa_{n}\frac{rB_{0}}{q\hat{\kappa}_{\perp}|\bm{\nabla}\psi|}\right)g_{1}(\vartheta),
    \end{equation}
  \end{subequations}
\end{widetext}
where $g_1(\vartheta)$ and $g_2(\vartheta)$ are the projections of $\Phi_s$ and $\delta P_\mathrm{comp}$ in ballooning space with convenient normalization, $\beta$ is the ratio of kinetic and magnetic pressure, $\omega_S$ is the sound frequency, $\kappa_g$ and $\kappa_n$ are geodesic and normal curvature, and $\hat{\bm{\kappa}}_\perp$ is the normalized perpendicular wave vector. The definitions of these quantities are as follows:
\begin{align*}
  &g_{1}(\vartheta)=\frac{\hat{\kappa}_\perp\hat{\Phi}_s(\vartheta)}{\left(\beta q^{2}\right)^{1/2}}\frac{ck_{\vartheta}}{B_0R_0},\quad
  g_{2}(\vartheta)=\frac{i\delta\hat{P}_\mathrm{comp}(\vartheta)}{\left(2\Gamma\right)^{1/2}P_0},\\
  &\beta=\frac{8\pi P_0}{B_0^2},\quad\omega_S^2=\frac{c_S^2}{q^2R_0^2},\quad\kappa_{g}=\frac{F(\psi)}{|\bm{\nabla}\psi|}\frac{\partial_{\theta}B_{0}}{\mathcal{J}B_{0}^{2}},\\
  &\kappa_{n}=\frac{|\bm{\nabla}\psi|}{B_{0}}\left(\partial_{\psi}B_{0}+\frac{4\pi}{B_{0}}\partial_{\psi}P_{0}\right)
  +\frac{\bm{\nabla}\psi\cdot\bm{\nabla}\theta}{|\bm{\nabla}\psi|}\frac{\partial_{\theta}B_{0}}{B_{0}},\\
  &\hat{\bm{\kappa}}_{\perp}=\frac{\bm{k}_{\perp}}{k_{\vartheta}}=
  s\vartheta\bm{\nabla}r+r\bm{\nabla}\theta-\frac{r}{q}\bm{\nabla}\zeta,
\end{align*}
where $k_\vartheta=-nq/r$, and $s=rq'/q$ is the magnetic shear. For simplicity, here, we have omitted the slow equilibrium radial scale dependence in the functions $g_1$ and g$_2$, noting that it is parameterized by the radial variation of equilibrium quantities. Eqs. \eqref{SAW-ISW}, describing the SAW-ISW mode structures propagating along the magnetic field lines, are the starting point of the present work. 

Eqs. \eqref{SAW-ISW} with proper boundary conditions define a well posed eigenvalue problem that generally doesn't allow for analytical solutions except in some particular cases with limiting parameters and simplified geometry \cite{FZoncaPoFB1993,LChenPoP2017}. In this work, we developed a workflow to solve Eqs. \eqref{SAW-ISW} numerically with proper boundary conditions. Compared with other codes that directly solve the ideal MHD equations in real space, adopting the ballooning mode representation in our model equations allows us not only to compute the high-n modes with higher resolution, but also to study the underlying physics in more detail, as will be shown below. In the following, we will briefly introduce our methodology for solving this problem, which was firstly proposed in Ref. \citenum{MFalessiPoP2019_continuum}.

\subsection{Solving method}\label{Solving method}

Taking the $|\vartheta|\rightarrow\infty$ limit, we can reduce Eqs. \eqref{SAW-ISW} to \cite{MFalessiPoP2019_continuum}
\begin{subequations}\label{SAW-ISW-limit}
  \begin{equation}
    \begin{aligned}
      &\left(\partial_{\vartheta}^2-\frac{\partial_{\vartheta}^2|\bm{\nabla} r|}{|\bm{\nabla} r|}+\frac{\omega^2\mathcal{J}^2B_0^2}{v_A^2}\right)g_1
      \\&=-\left(2\Gamma\beta\right)^{1/2}\frac{\mathcal{J}^2B_0^2}{qR_0}\frac{s\vartheta}{|s\vartheta|}\kappa_{g}g_2,
    \end{aligned}
  \end{equation}
   \begin{equation}
    \left(\frac{\mathcal{J}^{2}B_{0}^{2}}{q^{2}R_{0}^{2}}+\frac{\omega_{S}^{2}}{\omega^{2}}\partial_{\vartheta}^{2}\right)g_{2}
    =-\left(2\Gamma\beta\right)^{1/2}\frac{\mathcal{J}^2B_0^2}{qR_0}\frac{s\vartheta}{|s\vartheta|}\kappa_{g}g_1.
   \end{equation}
\end{subequations}
Notice that in Eq. \eqref{ballooning_representation}, due to finite magnetic shear, the extended poloidal angle $\vartheta$ also plays the role of a dimensionless radial wave vector. So Eqs. \eqref{SAW-ISW-limit}, obtained from Eqs. \eqref{SAW-ISW} in $|\vartheta|\rightarrow\infty$ limit, can actually describe the radial singular structures of plasma response, i.e., the continuous spectrum \cite{LChenPoP2017,MFalessiPoP2019_continuum,MFalessiJPP2020}. More generally, by solving Eqs. \eqref{SAW-ISW-limit}, we obtain the asymptotic behaviors of the parallel mode structures in the radial singular layer for both continuous spectrum as well as AEs \cite{FZoncaPoP2014a,FZoncaPoP2014b}. These solutions will then be taken as the boundary condition for Eqs. \eqref{SAW-ISW} to calculate parallel mode structure in the whole ballooning space and the corresponding dispersion relation. Here, we will consider the $\theta_k\equiv k_r/(nq') \rightarrow 0$ limit corresponding to the most unstable mode radial envelope structure, with $k_r$ the radial envelope wave number. At the same time, the radial envelope for the ground state of a radially localized mode structure can be computed analytically. Here, this will provide a simple ( "ad-hoc" ) example, which we will use to illustrate the two-dimensional radial mode structure that can be reconstructed following the present approach.

Following Ref. \citenum{MFalessiPoP2019_continuum}, we note that all the coefficients in Eqs. \eqref{SAW-ISW-limit}, related with the equilibrium quantities, are $2\pi$-periodic functions of $\vartheta$; so, Floquet theory \cite{GFloquet1883} can be employed to investigate the solutions of this system. In this paper, we review only the most important results of Floquet theory, and interested readers are referred to Refs. \citenum{GFloquet1883} and \citenum{MFalessiPoP2019_continuum} for more detailed discussions. Rewriting Eqs. \eqref{SAW-ISW-limit} as four first order coupled differential equations, we know from Floquet theory that, given the value of $\omega$, they must have solutions in the form of 
\begin{equation}\label{floquet_solutions}
  \mathbf{x}_i(\omega;\vartheta)=\mathbf{P}_i(\vartheta)e^{i\nu_i\vartheta},\quad i=1,2,3,4,
\end{equation}
where $\mathbf{x}_i$ is a column vector with four elements corresponding to the solutions of $g_1$, $\partial_{\vartheta}g_1$, $g_2$ and $\partial_{\vartheta}g_2$, respectively. $\mathbf{P}_i(\vartheta)$ is a $2\pi$-periodic function, and $\nu_i$ is generally a complex number called characteristic Floquet exponent. One of the major advantages of Floquet solutions is that their asymptotic behaviors are totally determined by the characteristic Floquet exponent $\nu_i$, so we can choose the solutions that satisfy the physical boundary condition (outgoing/decaying wave \cite{FZoncaPoP2014a,FZoncaPoP2014b} in $\vartheta$ space), with only the information of $\nu_i(\omega, r)$. Specifically, solutions in the continuum are traveling waves that propagate toward $|\vartheta|\rightarrow\infty$, so the Floquet exponent must satisfy $\mathrm{Im}\nu=0$ and $\mathrm{sgn}(\vartheta)\partial\mathrm{Re}\nu/\partial\omega>0$. Solutions in the frequency gap are standing waves which exponentially decay as $|\vartheta|\rightarrow\infty$, corresponding to AEs, so the Floquet exponent must satisfy $\mathrm{sgn}(\vartheta)\mathrm{Im}\nu>0$. The FALCON code, developed by Falessi et al, calculates the local dispersion curves $\nu(\omega, r)$ at each radial position and uses $\nu^2(\omega, r)=(nq(r)-m)^2$ to compute the global continuous spectrum \cite{MFalessiPoP2019_continuum}. The advantage and accuracy of this approach is due to the fact that the dispersion curves $\nu(\omega,r)$ are independent of $n$ and of any fast radial variation. In this work, we use the dispersion curves calculated by FALCON to determine the frequencies and parallel mode structures of AEs.

Since for generally up-down asymmetric equilibrium, the solutions of Eqs. \eqref{SAW-ISW} are asymmetric for parity transformation in $\vartheta$ space, we should calculate the boundary condition on the left ($\vartheta\rightarrow-\infty$) and right ($\vartheta\rightarrow+\infty$) separately. At each boundary, among the four solutions in Eq. \eqref{floquet_solutions}, generally only two of them satisfy the physical boundary conditions. So we have the general solutions of Eqs. \eqref{SAW-ISW-limit}, for negative and positive $\vartheta$, expressed by 
\begin{equation}\label{boundary_conditions}
  \begin{aligned}
  &\mathbf{x}^-(\omega;\vartheta)=w_1^-\mathbf{x}_1^-+w_2^-\mathbf{x}_2^-,\\
  &\mathbf{x}^+(\omega;\vartheta)=w_1^+\mathbf{x}_1^++w_2^+\mathbf{x}_2^+,
\end{aligned}
\end{equation}
respectively, where $\mathbf{x}_{1,2}^-$ and $\mathbf{x}_{1,2}^+$ are selected from $\mathbf{x}_i$ in Eq. \eqref{floquet_solutions} taking into account the boundary conditions, while $w_{1,2}^-$ and $w_{1,2}^+$ are weight coefficients of the linear combination. Since Eqs. \eqref{SAW-ISW-limit} is the limiting form of Eqs. \eqref{SAW-ISW} as $|\vartheta|\rightarrow\infty$, it is possible to connect the solutions of Eqs. \eqref{SAW-ISW}, denoted by $\mathbf{X}^-(\vartheta)$ (for negative $\vartheta$) and $\mathbf{X}^+(\vartheta)$ (for positive $\vartheta$), with the solutions of Eqs. \eqref{SAW-ISW-limit}: $\mathbf{X}^-(-2\pi\ell)=\mathbf{x}^-(\omega;0)$ and $\mathbf{X}^+(2\pi\ell)=\mathbf{x}^+(\omega;0)$, where $\ell$ is a sufficiently large positive integer. The solution in the whole ballooning space can be formally expressed by 
\begin{equation}\label{form_solutions}
  \begin{aligned}
  &\mathbf{X}^-(\vartheta)=w_1^-\mathbf{X}_1^-+w_2^-\mathbf{X}_2^-,\\
  &\mathbf{X}^+(\vartheta)=w_1^+\mathbf{X}_1^++w_2^+\mathbf{X}_2^+,
  \end{aligned}
\end{equation}
where $\mathbf{X}_{1,2}^-$ and $\mathbf{X}_{1,2}^+$ are obtained by numerically integrating Eqs. \eqref{SAW-ISW} from $-2\pi\ell$ and $2\pi\ell$ with boundary condition $\mathbf{X}_{1,2}^-(-2\pi\ell)=\mathbf{x}_{1,2}^-(\omega;0)$ and $\mathbf{X}_{1,2}^+(2\pi\ell)=\mathbf{x}_{1,2}^+(\omega;0)$, respectively. The two solutions given in Eqs. \eqref{form_solutions} must match at $\vartheta=0$, i.e., $\mathbf{X}^-(0)=\mathbf{X}^+(0)$, which yields a linear matrix equation 
\begin{equation}\label{matching}
  \mathbf{M}(\omega)\mathbf{w}=\mathbf{0},
\end{equation}
where $\mathbf{w}=(w_1^-,w_2^-,w_1^+,w_2^+)$ is a column vector composed of the four weight coefficients, and $\mathbf{M}$ is a $4\times4$ matrix given by 
\begin{equation}
  \begin{pmatrix}\mathbf{X}_1^-(0) & \mathbf{X}_2^-(0) & -\mathbf{X}_1^+(0) & -\mathbf{X}_2^+(0)
  \end{pmatrix}.
  \end{equation}
Condition for Eq. \eqref{matching} having nontrivial solution leads to the local dispersion relation $D(\omega)\equiv\mathrm{det}(\mathbf{M})=0$, where we have omitted its explicit dependence on the local parameters for simplicity. In practice, the eigenvalue $\omega$ is found by scanning the frequency gap and minimizing the value of $|D(\omega)|$. The corresponding solution of $\mathbf{w}$ is then substituted into Eqs. \eqref{form_solutions} to construct the parallel mode structure.

\section{Numerical results}\label{Numerical results}

\begin{figure}[]
  \centering
  \includegraphics[scale=0.9]{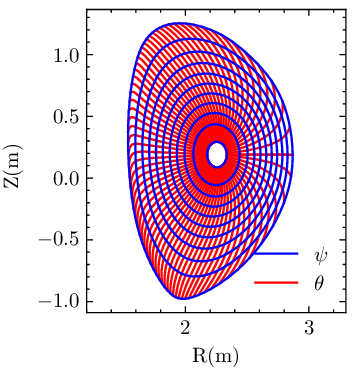}
  \caption{Contour lines of $\psi$ and $\theta$ coordinates.}\label{psi_theta} 
\end{figure}

\begin{figure}[]
  \centering
  \subfigure{
      \includegraphics[scale=0.5]{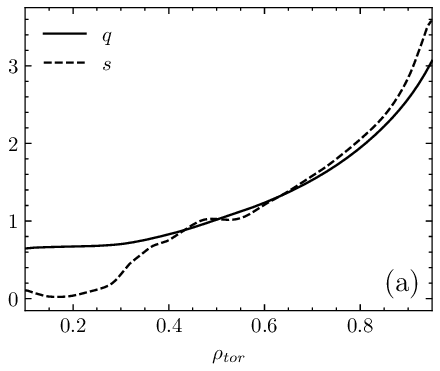}
  }
  \subfigure{
      \includegraphics[scale=0.5]{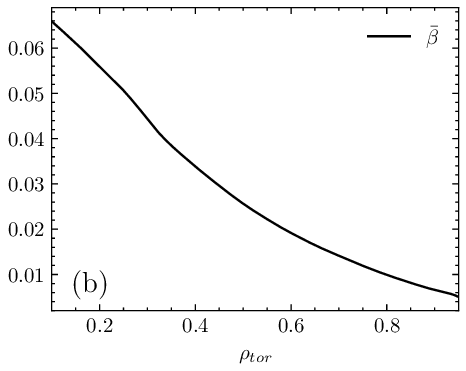}
  }
  \caption{(a) Profiles of safety factor $q$ and magnetic shear $s$. (b) Profile of $\bar{\beta}$. }\label{profiles}
\end{figure}

The model equations and numerical approach introduced in previous section for solving SAW-ISW eigenmodes is general. In this section, we will illustrate, taking the well-known TAE as an example, the calculation of mode frequency and parallel mode structure in general geometry of experimental interest. For consistency with Ref. \citenum{MFalessiPoP2019_continuum}, we use the equilibrium of a DTT reference scenario \cite{RAlbaneseFED2017} to illustrate the application of our code. The magnetic equilibrium has been originally calculated by means of the free boundary equilibrium evolution code CREATE-NL \cite{RAlbaneseFED2003} and further refined using the high-resolution equilibrium solver CHEASE \cite{HLutjensCPC1996}. We consider a single null configuration with magnetic axis and plasma edge excluded to avoid the possible singularity, as shown in FIG. \ref{psi_theta}. The profiles of safety factor and magnetic shear are depicted in FIG. \ref{profiles} (a), with $\rho_{tor}$ the normalized toroidal radius\cite{MFalessiPoP2019_continuum}. Profile of $\bar{\beta}\equiv 8\pi P_0/\bar{B}_0^2$ is depicted in FIG. \ref{profiles} (b), with $\bar{B}_0$ the magnetic field on axis.

\subsection{Boundary condition}\label{Boundary condition}

\begin{figure}[]
  \centering
  \subfigure{
      \includegraphics[scale=0.5]{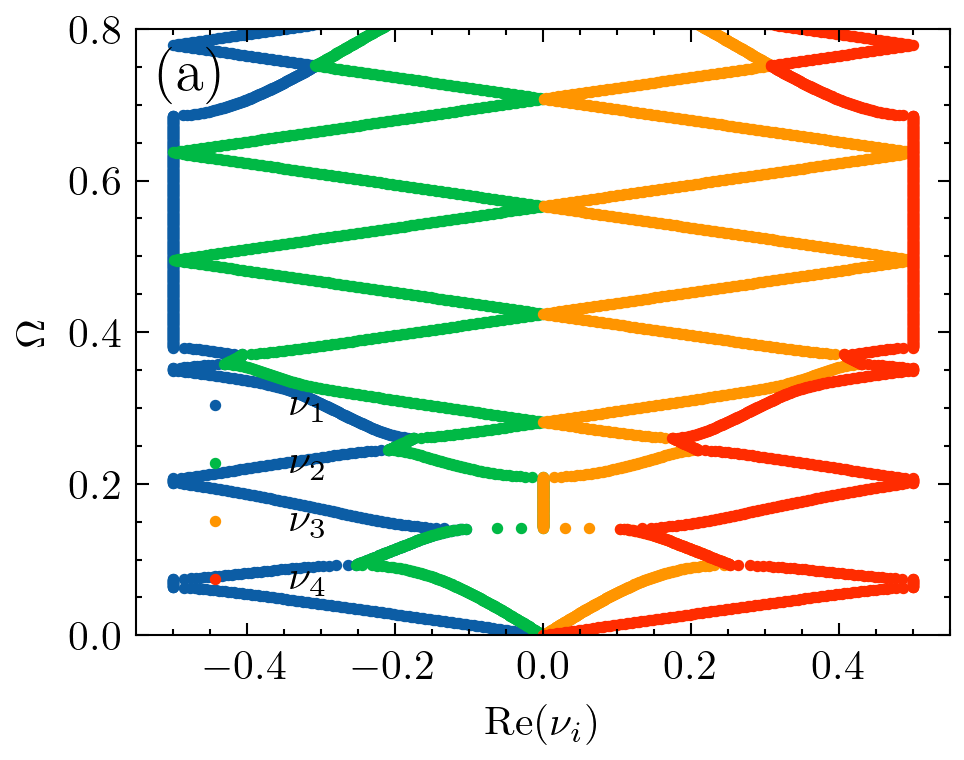}
  }
  \subfigure{
      \includegraphics[scale=0.5]{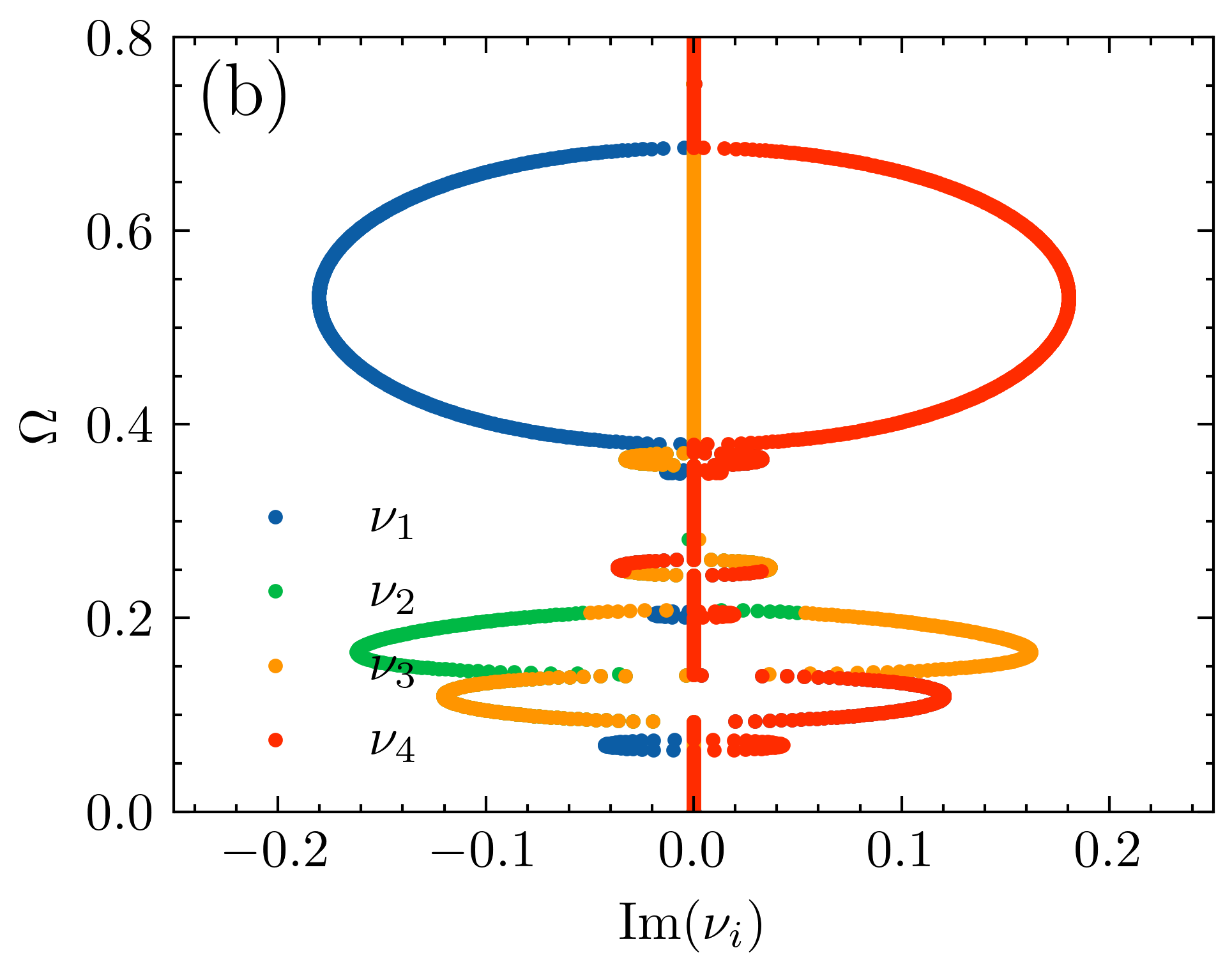}
  }
  \caption{(a) Real parts and (b) imaginary parts of the characteristic Floquet exponents $\nu_i(\Omega)$. }\label{local_dc}
\end{figure}

Following the workflow described in Sec. \ref{Solving method}, we firstly calculate all the coefficients of Eq. (6) using EQUIPE. Then we call FALCON to compute the dispersion curves and the boundary condition of Eqs. \eqref{SAW-ISW}. FIG. \ref{local_dc} (a) and (b) show the real and imaginary parts of $\nu_i(\Omega)$ at $\rho_{tor}=0.54$, respectively. Here, $\Omega=\omega R_0/\bar{v}_{A0}$ with $\bar{v}_{A0}$ the Alfvén velocity on the magnetic axis, and for convenience, $\Omega$ is chosen as the vertical coordinates to show different branches and gaps more clearly. Then we can pick out the SAW-ISW continuum with $\mathrm{Im}{\nu}=0$, as depicted in FIG. \ref{continuum_full} (a), where the shade of the dot's color is proportional to the value of Alfvénicity \cite{MFalessiPoP2019_continuum,MFalessiJPP2020,GPucellaPPCF2022}
\begin{equation}\label{alfvenicity}
\mathcal{A}=\frac{\int_0^{2\pi} g_1^2(\vartheta; \nu, r)d\vartheta}{\int_0^{2\pi} [g_1^2(\vartheta; \nu, r)+g_2^2(\vartheta; \nu, r)]d\vartheta},
\end{equation}
which characterizes the degree of SAW polarization of the fluctuation. Here, $g_1$ and $g_2$ denote the solutions of Eqs. \eqref{SAW-ISW-limit}. According to FIG. \ref{continuum_full} (a), there is clearly a SAW polarized frequency gap at $\Omega\sim0.5$, which is caused by the interaction of two counter-propagating SAWs along the magnetic field line. We know that TAE is located in this frequency gap, and the existence of acoustic polarized continuum in the same region implies the possible damping of TAE due to the coupling with ISW continuum, as will be shown in the next subsection. Some insights into the coupling of AEs and ISW continuum can be obtained by checking the properties of the characteristic Floquet exponents in FIG. \ref{local_dc} (a) and (b). Within the range $0.38\lesssim\Omega\lesssim 0.68$, $\nu_2=-\nu_3$ is purely a real number, which is related to the ISW continuum solution, while $\nu_1=-\nu_4$ has a real part of $-0.5$ and a negative imaginary part, which is related to the SAW gap AE solution. The asymptotic solution of TAE parallel mode structure in this frequency range is generally given by the mixture of both SAW gap and ISW continuum solutions 
\begin{equation}\label{boundary_conditions_TAE}
  \begin{aligned}
  &\bm{x}^-(\omega;\vartheta)=w_1^- P_1(\vartheta)e^{i\nu_1\vartheta}+w_2^- P_2(\vartheta)e^{i\nu_2\vartheta},\\
  &\bm{x}^+(\omega;\vartheta)=w_1^+ P_3(\vartheta)e^{i\nu_3\vartheta}+w_2^+ P_4(\vartheta)e^{i\nu_4\vartheta},
  \end{aligned}
\end{equation}
where outgoing wave boundary conditions have been imposed, and $\omega$ is assumed to be located at a position where $\partial\nu_2/\partial\omega=-\partial\nu_3/\partial\omega<0$, otherwise, subscripts $2$ and $3$ in Eqs. \eqref{boundary_conditions_TAE} should be exchanged.

\subsection{Frequency and parallel mode structure of TAE in DTT}

\begin{figure}[]
  \centering
  \subfigure{
    \includegraphics[scale=0.8]{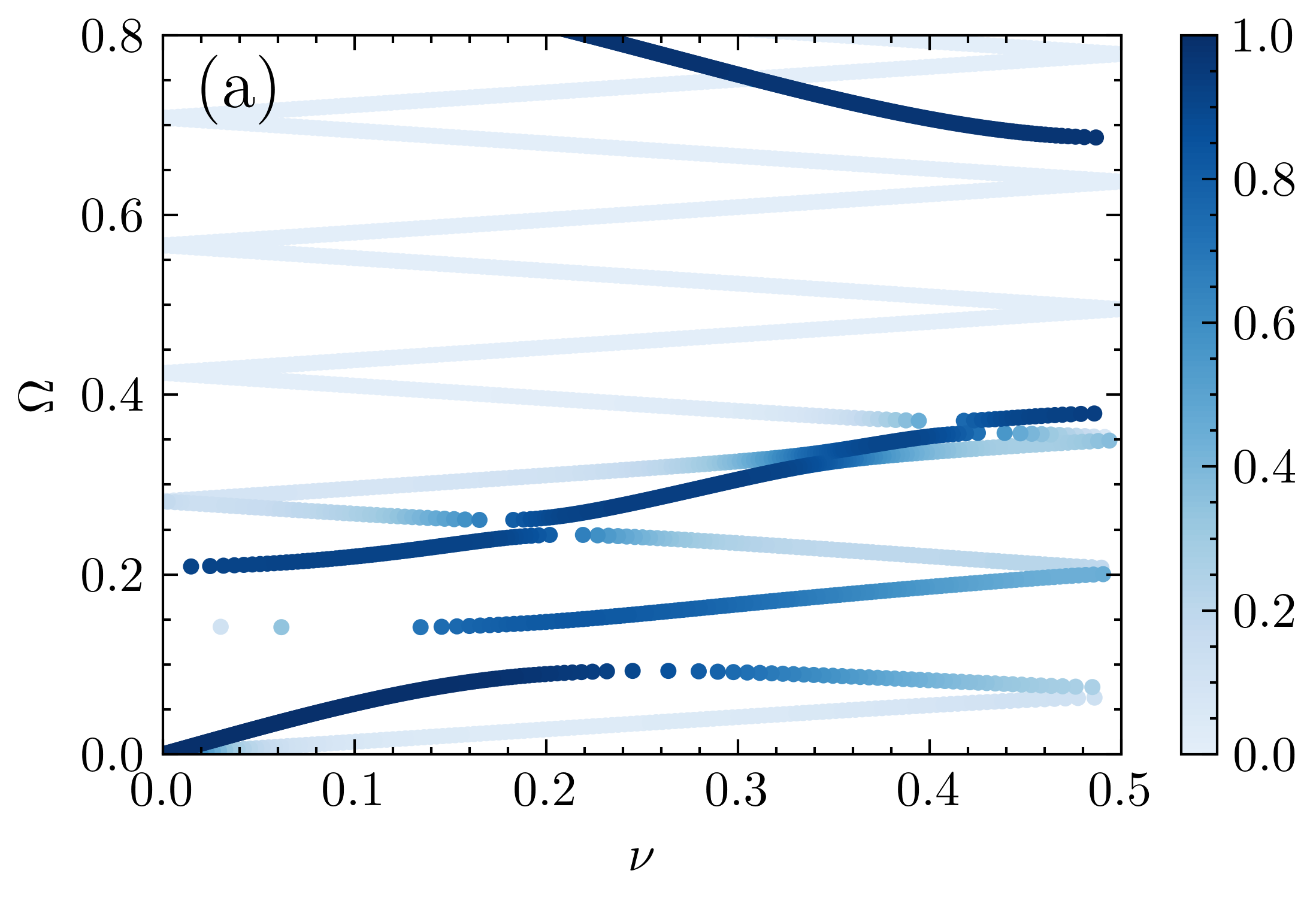}
  }
  \subfigure{
    \includegraphics[scale=0.8]{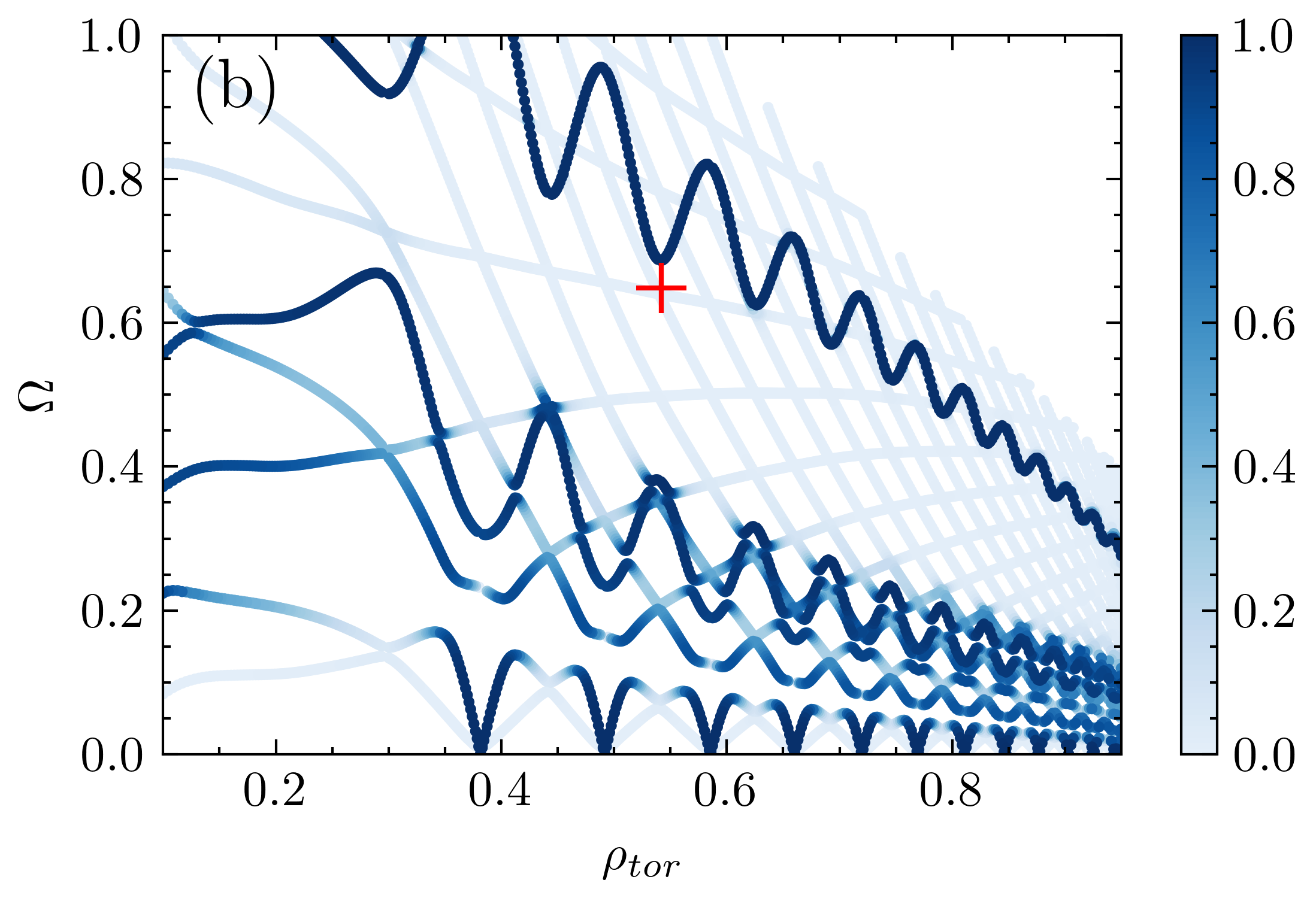}
  }
  \caption{(a) Local dispersion curves $\nu(\Omega)$ at $\rho_{tor}=0.54$, with the color-bar showing the Alfvénicity defined by Eq. \eqref{alfvenicity}, whose value is connected with the polarization of the corresponding fluctuation\cite{MFalessiPoP2019_continuum,MFalessiJPP2020}. (b) SAW-ISW continuous spectrum of $n=5$ calculated by FALCON in DTT, and the color-bar shows the Alfvénicity. The red plus "+" marks the frequency of TAE calculated at $\rho_{tor}=0.54$. }\label{continuum_full}
\end{figure}

In this subsection, we calculate the frequency and parallel mode structure of TAE in the DTT reference equilibrium. FIG. \ref{continuum_full} (b) depicts the SAW-ISW continuous spectrum of $n=5$ calculated by FALCON in this equilibrium, with Alfvénicity denoted by the color-bar. Given the boundary condition in Sec. \ref{Boundary condition}, and adopting the workflow introduced in Sec. \ref{Solving method}, we calculate the TAE frequency and parallel mode structure at the radial position $\rho_{tor}=0.54$, where $\bar{\beta}=2.28\%$. The real frequency of TAE is $\Omega_r=0.649$, close to the upper SAW continuum, as marked by the red plus "+" in FIG. \ref{continuum_full}. Meanwhile, coupling of TAE with the local ISW continuum will lead to a finite damping rate of the mode, due to the generation of short scale radial singular structure. Considering the different polarization of the AE and the resonantly excited ISW continuum (FIG. \ref{continuum_full} (a)), the damping rate of TAE $\Omega_i$ is expected to be very small. This justifies the use of a perturbation method to calculate $\Omega_i$. Upon determining the value of $\Omega_r$ by finding the roots of $D_r(\Omega_r)$, we then obtain the value of $\Omega_i$ as 
\begin{equation}\label{perturbed_expansion}
  \Omega_i=-\frac{D_i}{\partial D_r/\partial\Omega}\Big|_{\Omega=\Omega_r},
\end{equation}
where $D_r$ and $D_i$ are real and imaginary parts of $D(\Omega_r)$, respectively. In this way, the obtained value of $\Omega_i=-1.61\times10^{-4}$, which is much smaller than $\Omega_r$ as expected. 

\begin{figure}[]
  \centering
  \includegraphics[scale=0.8]{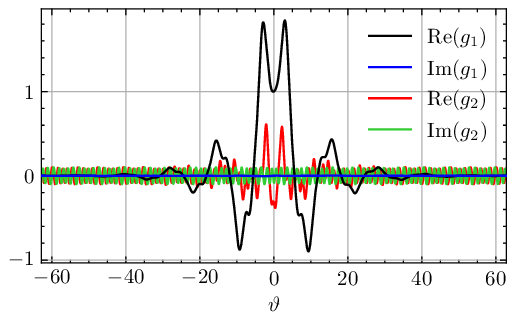}
  \caption{Normalized $g_1$ and $g_2$ in $\vartheta$ space.}\label{parallel_ms}
\end{figure}

\begin{figure}[]
  \centering
  \includegraphics[scale=0.8]{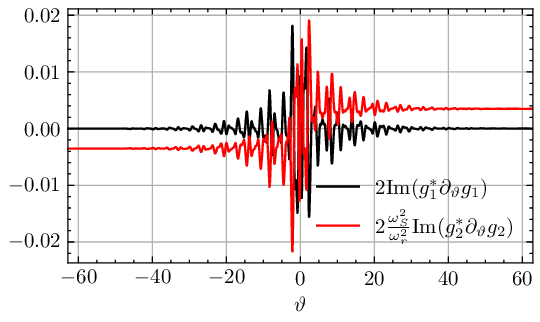}
  \caption{Energy fluxes in $\vartheta$ space. The black and red lines represents the energy flux contributed by 
  $g_1$ and $g_2$, respectively.}\label{energy_flux}
\end{figure}

The small but finite TAE coupling with ISW continuum and the consequent damping rate is consistent with the parallel mode structures of $g_1$ and $g_2$ in ballooning space, as shown in FIG. \ref{parallel_ms}. $g_1$, representing the response of SAW, is characterized by an exponentially decaying behavior corresponding the formation of a gap mode. Conversely, $g_2$, which represents the response of ISW, exhibits a periodic oscillating behavior typical of a continuum mode. The oscillation of $g_2$ is much faster than $g_1$, which can be explained by the fact that an Alfvén wave with relatively low $k_\parallel$ is interacting with sound waves with higher $k_\parallel$. The parallel mode structure in large-$\vartheta$ limit implies that there is an energy flux from large scale to small scale. This physical picture can be seen more clearly using a variational principle \cite{ZQiuPoP2012}. Specifically, we perform the following operation on Eqs. \eqref{SAW-ISW}
\begin{equation*}
  g_1^*\cdot\mathrm{(6a)}-g_1\cdot\mathrm{(6a)}^*+\left[g_2^*\cdot\mathrm{(6b)}-g_2\cdot\mathrm{(6b)}^*\right],
\end{equation*}
and integrate over the ballooning space. Thus, we get 
\begin{equation}\label{energy}
  \begin{aligned}
    &4i\omega_{r}\omega_{i}\int_{-\infty}^{\infty}\frac{\mathcal{J}^{2}B_{0}^{2}}{v_{A}^{2}}|g_{1}|^{2}d\vartheta\simeq
    -\left(g_1^*\partial_{\vartheta}g_1-g_1\partial_{\vartheta}g_1^*\right)\Big|_{-\infty}^{\infty}\\
    &-\frac{\omega_S^2}{\omega_r^2}\left(g_2^*\partial_{\vartheta}g_2-g_2\partial_{\vartheta}g_2^*\right)\Big|_{-\infty}^{\infty},
  \end{aligned}
\end{equation}
where "*" is the standard notation for complex conjugate. In Eq. \eqref{energy}, the left-hand side represents the changing rate of the total energy, and the two terms on the right-hand side represent the outgoing energy flux, contributed by $g_1$ and $g_2$, respectively. With the parallel mode structures of $g_1$ and $g_2$ already obtained numerically, we plot the energy fluxes of $g_1$ and $g_2$ in FIG. \ref{energy_flux}. The combination of Eq. \eqref{energy} and FIG. \ref{energy_flux} demonstrates that the energy of the mode primarily originates from $g_1$, whereas the energy flux is attributed to $g_2$, which is consistent with the physical picture that the TAE in the gap is slightly modified by the acoustic continuum. Furthermore, we calculate the damping rate directly given by Eq. \eqref{energy}, resulting in a value of $\Omega_i=-1.84\times10^{-4}$, which agrees well with the previous result from Eq. \eqref{perturbed_expansion}, obtained from a perturbation expansion. 

The damping rate of TAE resulting from the coupling with acoustic continuum is notably small, about $\mathcal{O}(10^{-4})$ of the real frequency. This damping rate can easily be covered by numerical errors in most codes. Here, the local analysis that we have adopted gives us the capability to investigate such a subtle physical effect. However, we should note that, the value of the damping rate is not always so small, as it depends on the strength of SAW-ISW coupling and, ultimately, on the mode polarizations \cite{MFalessiJPP2020}. It can be expected that, SAW and ISW may be more strongly coupled for higher $\bar{\beta}$, and the damping rate could be more relevant. For example, at $\rho_{tor}=0.385$, $\bar{\beta}=3.53\%$, the obtained TAE real frequency is $\Omega_r=0.770$, with a damping rate of $\Omega_i=-1.43\times 10^{-3}$. It is also worth mentioning that Eqs. \eqref{SAW-ISW} generally have multiple solutions satisfying the outgoing wave boundary conditions in the complex $\Omega$ space. The solution we have reported is the Alfvén polarized one, corresponding to the well-known TAE. Meanwhile, there also exist acoustic polarized solutions in the same frequency range, which are highly damped due to the strong coupling with the continuum and are, therefore, less relevant and ignored here. We emphasize, again, that mode polarization is the control physical quantity that determines how strongly damped the corresponding fluctuation structure is (cf. the color shade in Fig. \ref{continuum_full}, reflecting the Alfvénicity\cite{MFalessiPoP2019_continuum,MFalessiJPP2020,GPucellaPPCF2022}).

\begin{figure}[]
  \centering
  \includegraphics[scale=0.8]{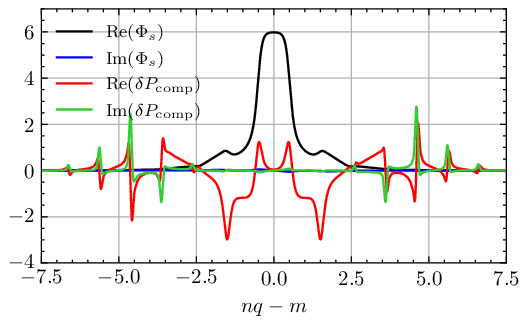}
  \caption{Mode structures of the $m$-th harmonics of the normalized $\Phi_s$ and $\delta P_{\mathrm{comp}}$ in real space. 
  The horizontal axis $nq-m$ can be regarded as the radial coordinate.}\label{real_space}
\end{figure}

For a more intuitive understanding of the physics mechanism underlying the TAE damping by resonant ISW continuum excitation, it is helpful to investigate the mode structures in real space. The $m$-th harmonics of the normalized $\Phi_s$ and $\delta P_{\mathrm{comp}}$ in real space, obtained by direct Fourier transform of $g_1/\hat{\kappa}_\perp$ and $g_2$ consistent with Eq. \eqref{ballooning_representation}, are shown in FIG. \ref{real_space}. There are clearly two dominant radial singular structures of $\delta P_{\mathrm{comp}}$ around $nq-m=\pm 4.5$, which reflect result of acoustic continuum resonances where $\omega^2-(nq-m)^2\omega_S^2=0$ is satisfied. Meanwhile, the sidebands around $nq-m=\pm 3.5, \pm 5.5$ arise from the coupling of different poloidal harmonics. By analogy with SAW resonant absorption\cite{LChenPoF1974,AHasegawaPRL1974}, one may expect the radial singular structures illustrated here bear acoustic continuum damping due to the same process. In the present case, the multiple resonance absorption layers must be attributed to the rich SAW-ISW coupling due to geometry effects\cite{MFalessiPoP2019_continuum, MFalessiJPP2020,GPucellaPPCF2022}. This makes the accurate calculation of the resonant absorption quite involved in the real space and shows the better accuracy of our present approach. Meanwhile, it is evident that the damping rate is relatively small since the resonances occur in high $|k_\parallel|$ region, where the dominant TAE mode structure ($\Phi_s$) rapidly decays away from its peak.

\subsection{Global mode structure of TAE in DTT} 
\begin{figure}[]
  \centering
  \subfigure{
      \includegraphics[scale=0.9]{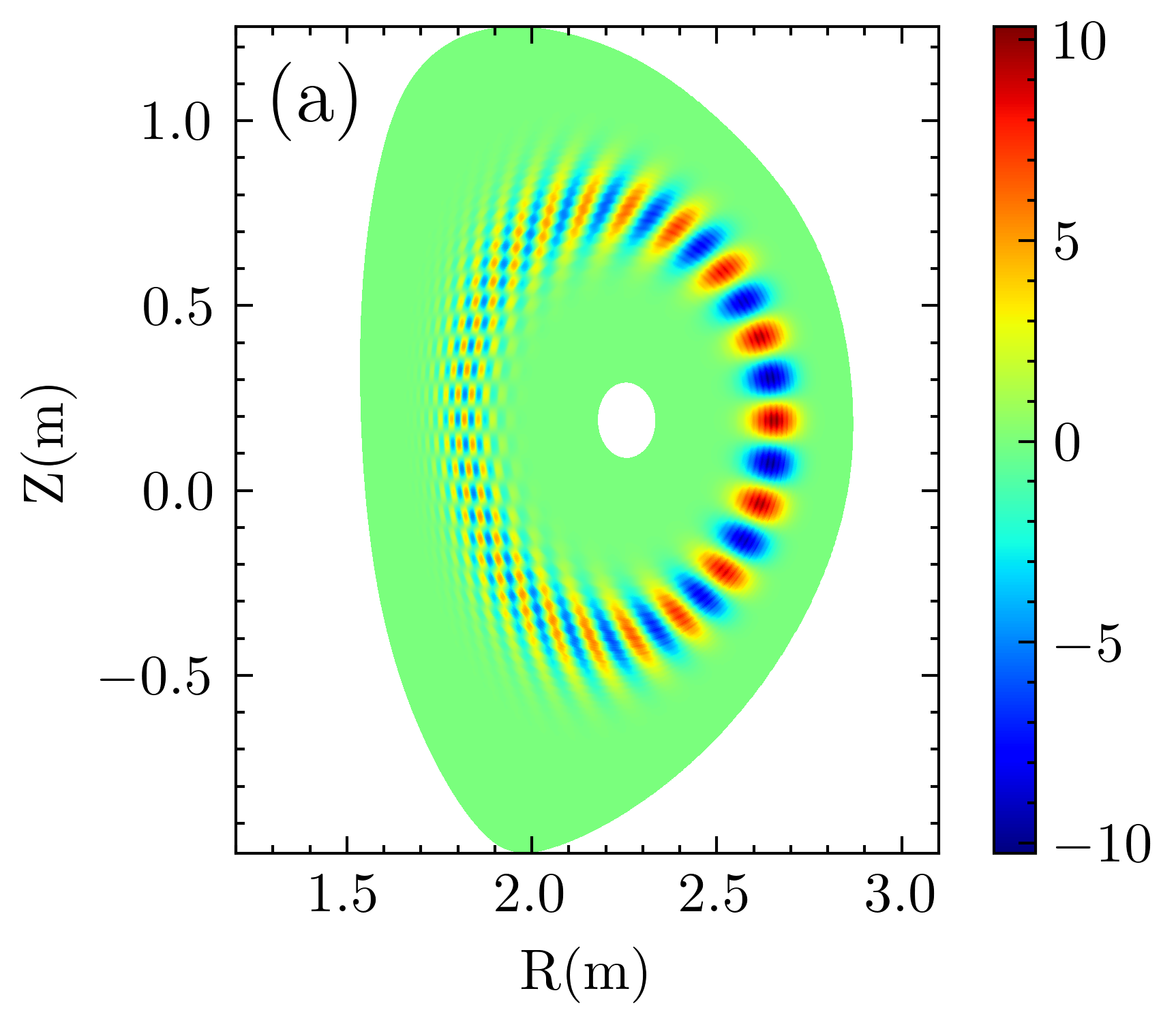}
  }
  \subfigure{
      \includegraphics[scale=0.2]{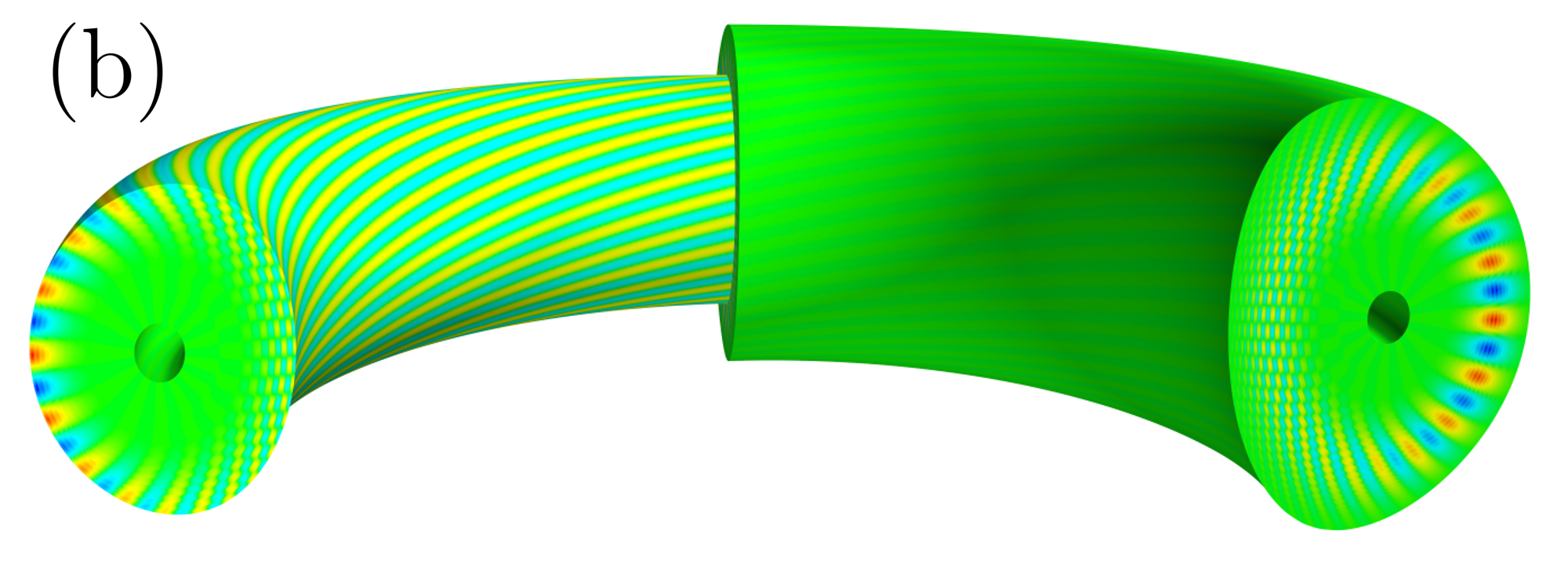}
  }
  \caption{(a) Two-dimensional and (b) three-dimensional mode structures of a $n=20$ TAE obtained with an "ad-hoc" 
  radial envelope.
  }\label{2d_3d_structures}
\end{figure}

In this study, we have omitted the slow radial variation of the parallel mode structures and focused on the solution of the local problem, leaving the self-consistent solution of the global problem to future studies. Nonetheless, as anticipated in Sec. \ref{Theoretical Framework},  the radial envelope for the ground state of a radially localized mode structure can be computed analytically \cite{FZoncaPoFB1993,YLiPPCF2023} and yields $A(r)=\exp[-(r-r_0)^2/\Delta r^2]$, where $r_0$ is the reference rational surface, and 
\begin{equation*}
\Delta r^2=\frac{2}{|nq'|}\left[\frac{\partial^2D/\partial\theta_k^2}{\partial^2D/\partial r^2}\right]
^{1/2}_{r=r_0}.
\end{equation*}
Here, $D(\omega,r,\theta_k)$ denotes the local TAE dispersion function. Using this analytical ("ad-hoc") radial envelope, we can schematically construct the global fluctuation fields in real space with the help of Eq. \eqref{ballooning_representation}. FIG. \ref{2d_3d_structures} (a) and FIG. \ref{2d_3d_structures} (b) show the two- and three-dimensional structures of the perturbed stream function $\Phi_s$ with toroidal mode number $n=20$. We note that FIG. \ref{2d_3d_structures} (a) is similar to the FIG. (7) in Ref. \citenum{YLiPPCF2023}, although a more accurate kinetic description is adopted in that work. Note that the radial singular structures due to SAW-ISW coupling are not visible in FIG. \ref{2d_3d_structures} due to the fact that the effect is finite but small, as discussed above. Actually, this radial singular will be completely eliminated in kinetic description, when the finite Larmor radius effect comes into play and modifies the potential functions \cite{YLiPPCF2023} in the model equations. Detailed studies of these physics effects will be reported elsewhere.

\begin{figure}[]
  \centering
  \includegraphics[scale=0.8]{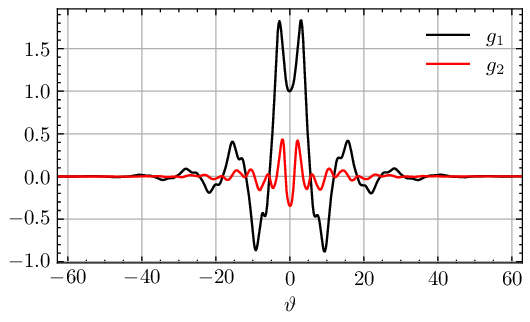}
  \caption{Normalized $g_1$ and $g_2$ in $\vartheta$ space, adopting slow sound approximation.}\label{parallel_ms_2d}
\end{figure}

As a final remark, it is worth mentioning that, for the high frequency modes such as TAE analyzed here, slow sound approximation (SSA) \cite{MSChuPoF1992,MFalessiPoP2019_continuum} is usually adopted to further simplify Eqs. \eqref{SAW-ISW}. The range of validity of this approximation has been systematically studied in Ref. \citenum{MFalessiJPP2020}. In SSA, one drops the $\partial_\vartheta^2$ term in Eq. \eqref{SAW-ISW-b} and turn it into an algebraic equation, so Eqs. \eqref{SAW-ISW} are reduced to one second order differential equation. By reapplying the methodology outlined in Sec. \ref{Solving method}, we calculate TAE in SSA using the same parameters. The frequency of TAE in SSA is $\Omega=0.645$, very close to the result in the full system. The corresponding parallel mode structure is shown in FIG. \ref{parallel_ms_2d}, which indicates that $g_1$ has little variation compared to that in the full system, while $g_2$ is significantly modified. This is indeed expected, since ISW resonances are removed in SSA, with the obvious consequence of removing TAE damping due to corresponding ISW resonance absorption. The comparison above between the results of SSA and the full system demonstrates that the former encapsulates most of the significant characters of TAE, such as the frequency and the parallel model structure of $g_1$, while the latter gives a finite amplitude of the parallel mode structure in singular layer and the resultant TAE damping rate. We note that, studying the fully compressible response will be crucial for the investigation of low-frequency AEs in future works, although rigorous calculation of the full system only introduces a perturbative correction compared to that of SSA for the high-frequency mode such as TAE analyzed here.

\subsection{Effect of NT on TAE}

\begin{figure}[]
  \centering
  \includegraphics[scale=0.9]{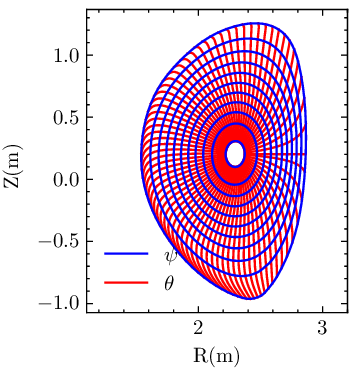}
  \caption{Contour lines of $\psi$ and $\theta$ coordinates in NT equilibrium.
  }\label{psi_theta_negative} 
\end{figure}

Recently, researches on using the NT plasma shaping as tokamak reactor design are revitalized due to its potential advantage in achieving a better plasma confinement while avoiding the threat of edge localized modes (ELMs) \cite{YCamenenNF2007,MAustinPRL2019,ANelsonPRL2023}. However, present researches on NT have not yet fully explored the spectrum of $n\sim\mathcal{O}(10)$ AEs \cite{TWangPoP2018}, whose impact on alpha particle transport could play a crucial role in the burning plasma power balance. In order to study the effect of different triangularities on TAEs, we calculate TAE in one of the proposed DTT NT quilibria and compare the results with those obtained previously in PT equilibrium. The cross section of the NT equilibrium is depicted in FIG. (\ref{psi_theta_negative}), which is obtained by mirror flipping the last magnetic surface in FIG. (\ref{psi_theta}) and recalculating the equilibrium using CHEASE code at fixed $q$ and pressure profiles, enabling us to isolate and analyze the specific impact of triangularity.

\begin{figure}[]
  \centering
  \includegraphics[scale=0.8]{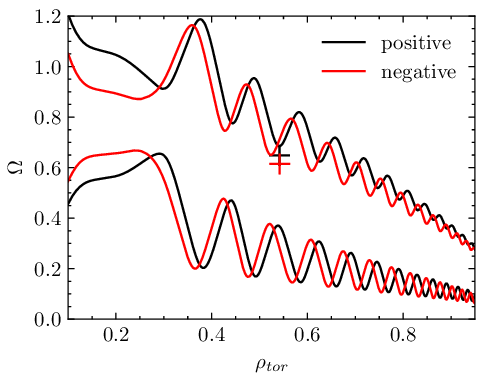}
  \caption{Continuous spectra of PT and NT equilibria, adopting slow sound approximation.
  }\label{continuum_slow} 
\end{figure}

\begin{figure}[]
  \centering
  \includegraphics[scale=0.8]{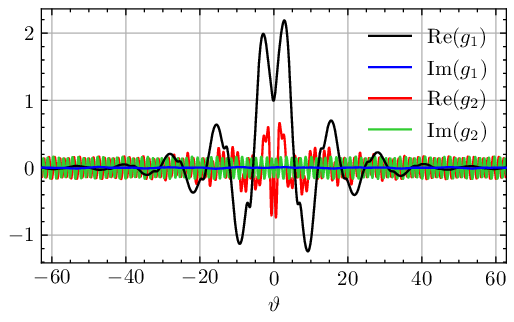}
  \caption{Normalized $g_1$ and $g_2$ in $\vartheta$ space in NT equilibrium.}\label{parallel_ms_negative}
\end{figure}

FIG. (\ref{continuum_slow}) depicts the continuous spectra of PT and NT equilibria calculated by EQUIPE-FALCON, where the slow sound approximation is adopted to focus on the variation of the TAE gap. The lower branches of the continuous spectra in both equilibria are situated within the same frequency range, while the upper branch is slightly lower in the NT equilibrium. We calculate the TAE frequency in this NT equilibrium, which is found to be $\Omega=0.616-1.67\times10^{-4}i$, and also slightly lower than that of PT equilibrium ($\Omega=0.649-1.61\times10^{-4}i$). The real frequencies of TAEs in these two equilibria are marked by black and red pluses in FIG. \ref{continuum_slow}, respectively, and their relative difference is negligibly small. The parallel mode structure of TAE is depicted in FIG. \ref{parallel_ms_negative}, which, as expected, shows a similar pattern with before (FIG. \ref{parallel_ms}). Therefore, we conclude that, in the ideal MHD limit, simply changing the triangularity of the equilibrium does not have a significant influence on the frequency or mode structure of TAE. This result is not surprising, considering that TAE spectrum has a predominant feature at $|\mathcal{J} B_0 k_\parallel | \sim 1/2$. Thus, the TAE dispersion relation, written in the form of a variational principle\cite{FZoncaPoP2014a,FZoncaPoP2014b}, readily shows that triangularity effects average out at the leading order. This is also consistent with toroidal frequency gap formation being dominanted by the $\cos\theta$-type variation of the magnetic field (i.e., toroidicity), while triangularity merely introduces a small component of higher order sidebands. In addition, since TAEs are radially localized deeper in the core region as compared to ELMs, triangularity may also attenuate its effect as one moves from plasma edge towards the magnetic axis. Our present results are consistent with other analyses reported in the literatures\cite{YGhaiNF2021,POyolaIAEA2023} about the relatively weak impact of NT on AE fluctuation structures. They also confirm that NT could instead be relevant in the AE excitation mechanism by EP via modification of both resonance conditions and EP orbits\cite{YGhaiNF2021,POyolaIAEA2023,FZoncaNJP2015}. Adopting the workflow proposed here, a more detailed investigation on this issue with kinetic effects included will be reported in our future work.

\section{Summary and prospect}\label{Summary and prospect}
In this work, we illustrate the numerical calculation of AEs in the ideal MHD limit, adopting the general theoretical framework discussed in Refs. \citenum{FZoncaPoP2014a}, \citenum{FZoncaPoP2014b}, \citenum{MFalessiPoP2019_continuum} and \citenum{MFalessiJPP2020}. By solving the model equations in the ballooning space, the asymptotic behaviors of the solutions in the large-$|\vartheta|$ limit are completely determined by the corresponding characteristic Floquet exponents, which makes it straightforward to select the proper boundary conditions and, thus, defines a well posed eigenvalue problem in the ballooning space. To solve this problem efficiently, a workflow has been developed on the basis of FALCON\cite{MFalessiPoP2019_continuum, MFalessiJPP2020}, which can be used to calculate the frequencies and parallel mode structures of AEs in realistic tokamak geometries. As an application of the code, we calculate the TAE in the reference equilibria of DTT and demonstrate it has a small but finite damping rate due to the coupling with the acoustic continuum. The existence of the damping rate is also confirmed by the derived variational principle, which highlights that the acoustic polarized component of the fluctuation carries a finite amount of energy leaking out of the potential well in ballooning space; i.e., toward radial singular structures, resulting in the dissipation of the mode. Although the obtained TAE damping rate for the parameters used here is small, due to the different polarization of the eigenmode and the ISW continuum, one should note that, in fusion plasma with higher $\beta$, SAW and ISW may be more strongly coupled even in the TAE frequency range, and the mechanism that we have proposed could potentially become more relevant\cite{MFalessiJPP2020}. By comparing the numerical results in different DTT equilibria with PT and NT, we find that simply changing the triangularity does not have a significant influence on TAE frequency and mode structure within the ideal MHD description. 

Although we have limited ourselves to the local problem by neglecting the slow variation of the radial envelope in Eqs. \eqref{SAW-ISW} for the sake of simplicity, the self-consistent solution of the global problem is included in the general theoretical framework \cite{FZoncaPoP2014a,FZoncaPoP2014b}. The general approach adopted in the present work can be straightforwardly extended to kinetic description due to the similar structure of the mode equations\cite{FZoncaPoP2014b}, allowing to explore the effects of finite Larmor radius, wave-particle interaction and diamagnetic frequency on the SAW-ISW fluctuations, as anticipated in Ref. \citenum{YLiPPCF2023}.

\section*{Acknowledgements}
The authors acknowledge Prof. Liu Chen (Zhejiang University, China) for valuable discussions. This work was supported by the National Science Foundation of China under Grant Nos. 12275236 and 12261131622, and Italian Ministry for Foreign Affairs and International Cooperation Project under Grant No. CN23GR02. This work was also supported by the EUROfusion Consortium, funded by the European Union via the Euratom Research and Training Programme (Grant Agreement No. 101052200 EUROfusion). The views and opinions expressed are, however, those of the author(s) only and do not necessarily reflect those of the European Union or the European Commission. Neither the European Union nor the European Commission can be held responsible for them.


%

\end{document}